\pdfoutput=1
\documentclass[prl,showpacs,twocolumn]{revtex4-1}
\usepackage{graphicx}
\usepackage{amsmath}
\usepackage{amssymb}
\usepackage{xspace}
\usepackage{pdfpages}
\usepackage{amsfonts}

\begin{document}
\title{Evidence for a Quantum-to-Classical Transition in a Pair of Coupled Quantum Rotors}
\author{Bryce Gadway}
\email{present address: JILA, University of Colorado, 440 UCB, Boulder, CO 80309-0440}
\author{Jeremy Reeves}
\author{Ludwig Krinner}
\author{Dominik Schneble}
\affiliation{Department of Physics and Astronomy, Stony Brook University, Stony Brook, NY 11794-3800, USA}
\date{\today}
\begin{abstract}
The understanding of how classical dynamics can emerge in closed quantum systems is a problem of fundamental importance. Remarkably, while classical behavior usually arises from coupling to thermal fluctuations or random spectral noise, it may also be an innate property of certain isolated, periodically driven quantum systems. Here, we experimentally realize the simplest such system, consisting of two coupled, kicked quantum rotors, by subjecting a coherent atomic matter wave to two periodically pulsed, incommensurate optical lattices. Momentum transport in this system is found to be radically different from that in a single kicked rotor, with a breakdown of dynamical localization and the emergence of classical diffusion. Our observation, which confirms a long-standing prediction for many-dimensional quantum-chaotic systems, sheds new light on the quantum-classical correspondence.
\end{abstract}
\maketitle

In ultracold atomic systems, the quantum nature of matter can be made manifest in striking ways. One example arises in the dynamics of quantum chaotic systems~\cite{Haake}, i.e. in  systems whose classical counterparts are chaotic and in which destructive interference can suppress the onset of chaos. A paradigm model, the $\delta$-kicked rotor~($\delta$-KR), can been realized with atomic matter waves subject to a periodically pulsed optical lattice~\cite{Moore-Qdeltakicked-1995}. Whereas regimes of fully chaotic behavior with diffusive growth of the momentum variable are expected in the classical case, destructive interference leads to dynamical localization~\cite{Casati-Israilev-1979,HoggHuberman-Recurrences-1982,Fishman-Chaos-Recurr-Anderson-1982} for which the momentum distribution remains frozen.  The phenomenon of dynamical localization in the $\delta$-KR is a direct analog of real-space Anderson localization in one-dimensional (1D) disordered materials~\cite{Grempel-quantumNonintegrable-1984}, and recent experimental work ~\cite{Ringot-Delocalization-QuasiPeriodic,Chabe-AndersonMetal-2008,Lemarie-AndersonTheoryExpt-2009,Lemarie-CriticalStateAnderson-2010} based on a generalization to quasi-periodic kicking~\cite{Casati-3Danderson-1989} has also provided access to the three-dimensional case.

\begin{figure}[hb!]
    \centering
    \includegraphics[width=3.15in]{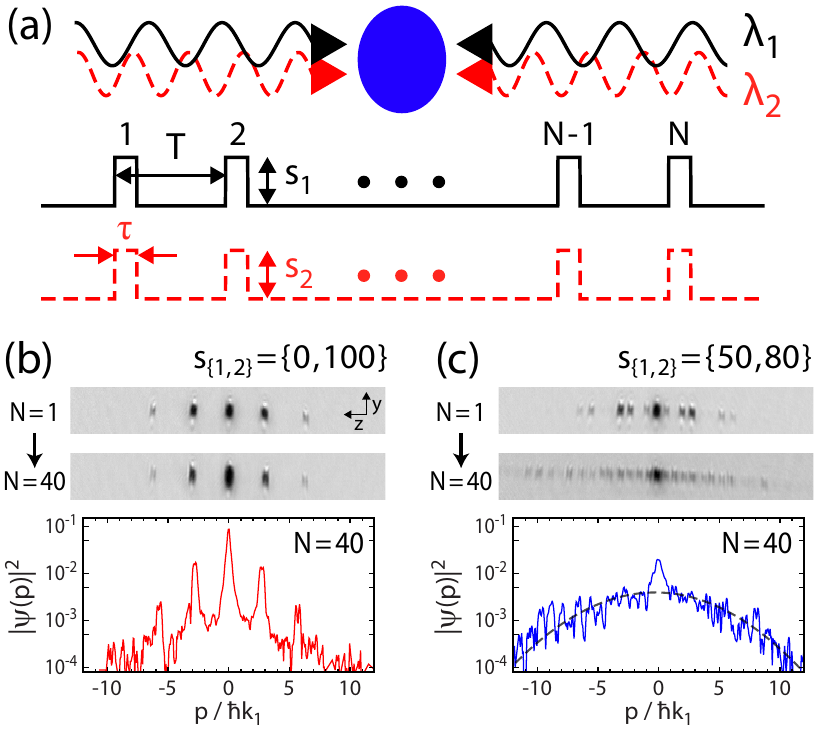}
\caption{Atomic matter waves in a periodically pulsed optical lattice potential.
(a)~A Bose--Einstein condensate is exposed to a train of $N$ pulses (duration $\tau$, separation $T$) of two incommensurate optical lattices (wavelengths $\lambda_{1,2}$ and depths $s_{ \{1,2\} }$).
(b)~Time-of-flight diffraction spectra (averaged over 3-4 images) of atoms released after $N=1$ and 40 kicks, for driving with a single lattice ($s_{ \{1,2\} } = \{0, 100\}$). The momentum distribution along $z$ (integrated along $y$) after $N=40$ kicks is shown in the bottom.
(c)~As in (b), but for driving with two incommensurate lattices ($s_{ \{1,2\} } = \{50,80\}$). The dashed black line at the bottom of (c) is a Gaussian profile corresponding to diffusive spreading.
}
    \label{FIG:Cartoon}
\end{figure}

Several experimental studies  of the $\delta$-KR model ~\cite{Ammann-98,Klappauf-NoiseDiss-DynLoc-1998,dArcy-QuantDiffusion-DKR-2001,dArcy-QaccelModes-Decoherence-2001,Zhang-Raizen-Interactions-2004,Duffy-NonlinearDKR-2004} have focused on the degradation of dynamical localization in the presence of noise \cite{Shepelyansky-Physica-1983,Ott-NoiseQuantChaos-1984,Casati-3Danderson-1989,Kolovsky-2color-1996} and nonlinearities \cite{Adachi-QuantToClass-CoupledRotors,Benvenuto-Nonlinear-1991,Shepel-Nonlinear-1993}.
Remarkably, signatures of classical behavior have been predicted ~\cite{Adachi-QuantToClass-CoupledRotors} to emerge already in a simple driven quantum system consisting of just two coupled kicked rotors, providing hope that the disparate behavior of quantum and classically chaotic systems may be reconciled in the macroscopic limit.

In this paper we realize such a simple coupled quantum system by subjecting a macroscopic matter wave to two periodically pulsed, incommensurate optical lattices~\cite{RoatiIncLatt}. As detailed further below, the coupling between the two rotors, each separately driven by one of the lattices, arises from the kinetic evolution between the pulses. We find that the coupling destroys hallmark behavior of the off-resonantly driven $\delta$-KR system, causing a transition from dynamical localization to classically diffusive momentum-space transport. Additionally, we observe that the coupling greatly modifies the response of atoms to resonant driving, leading to a suppression of ballistic transport in momentum space.

Our system consists of an optically-trapped Bose-Einstein condensate of ($1.4 \pm 0.4) \times 10^5$ $^{87}$Rb atoms in the $|F,m_F\rangle=|2,-2\rangle$ hyperfine ground state, which is subject to two simultaneously pulsed, incommensurate optical lattices~\cite{RoatiIncLatt} along $z$, as depicted in Fig.~\ref{FIG:Cartoon}~(a). The lattices have wavelengths $\lambda_1 = 1064$~nm and $\lambda_2=782$~nm (wave numbers $k_{1(2)} = 2\pi / \lambda_{1(2)}$) and lattice depths $s_{1(2)} E_R$, where $E_R = \hbar^2 k_1^2 / 2 M$ is the recoil energy of the first lattice and $M$ the atomic mass. The pulses have a duration $\tau = 2$~$\mu$s (Raman--Nath regime) and are spaced at a variable period $T$. After applying $N$ pulses, we immediately release the atoms and allow them to freely evolve in time-of-flight for $16$~ms before performing absorptive imaging of momentum distributions, with examples shown in Figs.~\ref{FIG:Cartoon}~(b,c).

Ignoring effects of the trapping potential and atom-atom interactions, this system can be approximately described by the 1D Hamiltonian $H = - \hbar^2 \partial_z^2  /2M + S(z) \sum_{j=1}^{N} \sqcap (t/\tau - jT/\tau)$, where $S(z) = [s_1 \cos(2k_1 z) + s_2 \cos(2 \eta k_1 z)]E_R/2$, $\sqcap$ is the normalized boxcar function, and $\eta = k_2 / k_1 \sim 1.36$ is the ratio of wavenumbers. Diffraction by the two optical lattices connects the zero-momentum condensate to modes with momenta in multiples of $2 \hbar k_1$ and $2 \eta \hbar k_1$, respectively. The two sets of modes have no intersection for irrational values of $\eta$, which allows us to describe the system as effectively 2D in the plane wave basis $|m,n\rangle = |m\rangle \bigotimes |n\rangle$, with momenta $p_{m,n} = 2(m + \eta n)\hbar k_1$  (we assume that the mode separation exceeds the spectral width of the condensate).

Approximating the lattice pulses as $\delta$-functions, the effective 2D Hamiltonian for the system is given by
\begin{equation}
H = H_T + \hbar (\hat{\phi}_{V_1} + \hat{\phi}_{V_2}) \sum_{j=1}^{N} \delta (t - j T) \ ,
\label{EQ:Hamm1}
\end{equation}
where $H_T = \hbar / T \sum_{m,n} D_{m,n} \hat{n}_{m,n}$ describes the kinetic energy of the plane-wave modes with
\begin{equation}
D_{m,n} = \kappa ( m^2 /2 + \eta m n + \eta^2 n^2 / 2 ) \ ,
\label{EQ:Hamm2}
\end{equation}
and the effect of the pulsed optical potential is captured by
\begin{equation}
\hat{\phi}_{V_{1(2)}} = (K_{1(2)} / \kappa) \sum_{m,n} (\hat{\sigma}^{-,1(2)}_{m,n} + \hat{\sigma}^{+,1(2)}_{m,n}) \ ,
\label{EQ:Hamm3}
\end{equation}
where $\hat{\sigma}^{\pm,1}_{m,n} = \hat{a}^{\dag}_{m \pm 1,n} \hat{a}_{m,n}$ and $\hat{\sigma}^{\pm,2}_{m,n} = \hat{a}^{\dag}_{m,n \pm 1}\hat{a}_{m,n}$ describe transitions within each set of modes.  Here, $\hat{a}^{\dag}_{m,n}$ ($\hat{a}_{m,n}$) is the creation (annihilation) operator and $\hat{n}_{m,n}$ is the number operator of the composite mode $|m,n\rangle$.
As in the standard treatment of the $\delta$-KR~\cite{Moore-Qdeltakicked-1995}, which is realized when either lattice is pulsed alone, we define $\kappa=8 E_R T / \hbar$ and $K_{1(2)} = \kappa s_{1(2)} E_R \tau/ 2 \hbar$. Here, the single-rotors are driven resonantly whenever $\kappa / 4\pi$ ($\eta^2 \kappa / 4\pi$) is a rational number, i.e. whenever the frequency of $\delta$-kicking matches a Talbot resonance~\cite{Deng-Talbot-1999,Ryu-HighOrderRes-2006}, and the stochasticity parameters $K_{1(2)}$ delineate regimes of regular and chaotic motion in the classical $\delta$-KR model.

\begin{figure}[b]
    \centering
    \includegraphics[width=3.15in]{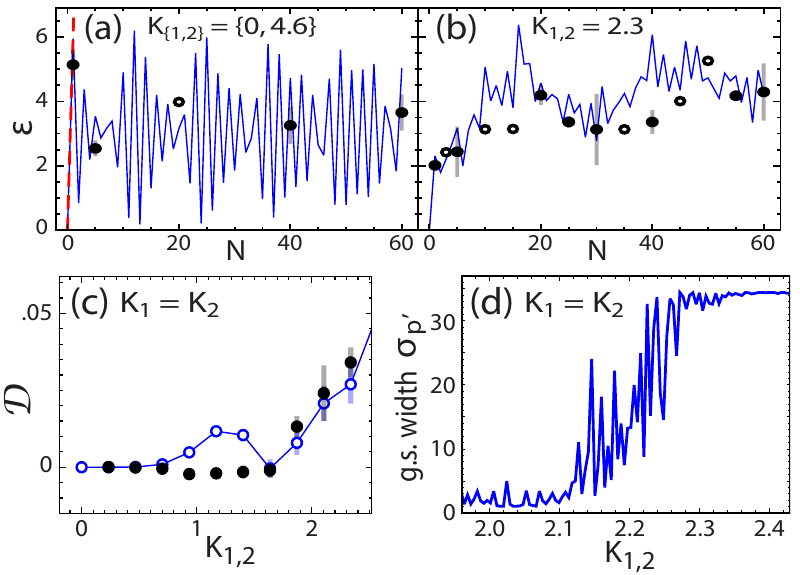}
\caption{Matter-wave dynamics for off-resonant kicking.
(a)~Kicking by a single lattice, with $K_2 = 4.6$. Shown is the atoms' per-particle energy $\varepsilon$ as a function of kick number $N$. The black points are experimental data with statistical error bars (empty circles for individual runs) while the blue solid and red dashed lines are simulated quantum and classical trajectories.
(b)~Off-resonant kicking with two incommensurate lattices ($K_1 = K_2 = 2.3$).
(c)~Energy diffusion rates, $\mathcal{D} = \Delta \varepsilon / \Delta N$, as determined from linear fits to data and simulated points as in (b), as a function of the kicking strength for the uniform case $K_1 = K_2 \equiv K_{1,2}$. Filled black points are experimental, open blue disks represent numerical, and error bars represent the standard error of the linear fits.
(d)~Calculated rms momentum width $\sigma_{p'}$ (in units $\hbar k_1$) for the ground state of the single-kick Floquet operator $\hat{U}$.
}
    \label{FIG:OffRes1}
\end{figure}

A simple picture of the $\delta$-KR and its connection to the Anderson model emerges from a stroboscopic Floquet analysis~\cite{Grempel-quantumNonintegrable-1984}, in which the effect of each kick is described by the operator $\hat{U} = \exp{[-i(\hat{\phi}_{V_1} + \hat{\phi}_{V_2})]} \exp{[-i \sum_{m,n} D_{m,n} \hat{n}_{m,n}]}$, such that the initial state $|\psi_{0}\rangle = |m=0,n=0\rangle$ is transformed to  $|\psi_{N}\rangle = \hat{U}^N |\psi_{0}\rangle$ after a series of $N$ kicks. The term $D_{m,n}$ (taken modulo $2\pi$) contained in the operator $\hat{U}$ describes the kinetic phase evolution between kicks. In the language of the Anderson model, it represents a 2D quasienergy landscape, within which the terms $K_{1(2)}/\kappa$ control the strength of discrete-time hopping.

We first discuss a single rotor at quantum resonance, which is obtained e.g. for the first lattice when $\kappa / 4\pi$ is a rational number. In this case, the relevant quasienergy term from Eq.~\ref{EQ:Hamm2} is equal to zero (mod $2\pi$) for all of the momentum orders, corresponding to a flat quasienergy landscape in which tunneling occurs. Thus, atoms initially localized to a single momentum mode will undergo ballistic momentum-space transport~\cite{Deng-Talbot-1999,Ryu-HighOrderRes-2006} corresponding to a quantum walk~\cite{Romanelli-GeneralizeQRW-2005,*Romanelli-QKR-QRW-2007}. Contrastingly, if the resonance condition is not fulfilled, the quasienergy landscape is characterized by pseudorandom disorder in which the atomic wavefunction dynamically localizes~\cite{Moore-Qdeltakicked-1995}.

In our case of two kicked rotors, the 2D quasienergy landscape is anisotropic. The quantities $\kappa/2$ and $\eta^2 \kappa/2$ control the disorder strengths in the two directions, and in the absence of the coupling, this 2D system should display localization for any finite disorder strength~\cite{Abrahams-Anderson2D-1979}. However, in our system a coupling between the rotors arises because the geometry remains physically one-dimensional and because the free-space dispersion relation for massive particles is quadractic. In the following, we shall investigate the effect of the coupling term $\kappa \eta m n$ in Eq.~\ref{EQ:Hamm2} on the localization properties of the system.

In our experiment, we first tune the pulse period $T$ such that the kicking is off-resonant for both lattices ($T=36~\mu$s; $\kappa / 4\pi \approx 0.29$~;~$\eta^2\kappa / 4\pi \approx 0.54$).
In the single-lattice case, the atomic population remains trapped in the lowest momentum orders and the per-particle energy $\varepsilon$ (in units of $E_R$) shows no net increase over a large number of pulses, as shown in Fig.~\ref{FIG:Cartoon}~(b) and Fig.~\ref{FIG:OffRes1}~(a) for $K_{ \{1,2\} } = \{0,4.6\}$, in agreement with the expectation for dynamical localization, which occurs after a ``quantum break time'' $t_B = T K^2 / 4 \kappa^2$~\cite{Moore-Qdeltakicked-1995,Klappauf-NoiseDiss-DynLoc-1998}. For our system parameters, $t_B$ is smaller than $T$ (similar as in \cite{Duffy-NonlinearDKR-2004}), such that dynamical localization sets in immediately.
This finding is confirmed by an exact numerical simulation in the plane wave basis of states $|m,n\rangle$ with momentum $p_{m,n} = 2(m + \eta n)\hbar k_1$ which displays a fast oscillatory behavior around a constant mean energy~\cite{DKR-SuppMatNEW}. The lack of growth is also in stark contrast to computed classical trajectories, whose energy increases very rapidly, cf. Fig.~\ref{FIG:OffRes1}~(a), ruling out effects of classical localization~\cite{Geisel-KAM-1986}.

\begin{figure}[b]
    \centering
    \includegraphics[width=3.15in]{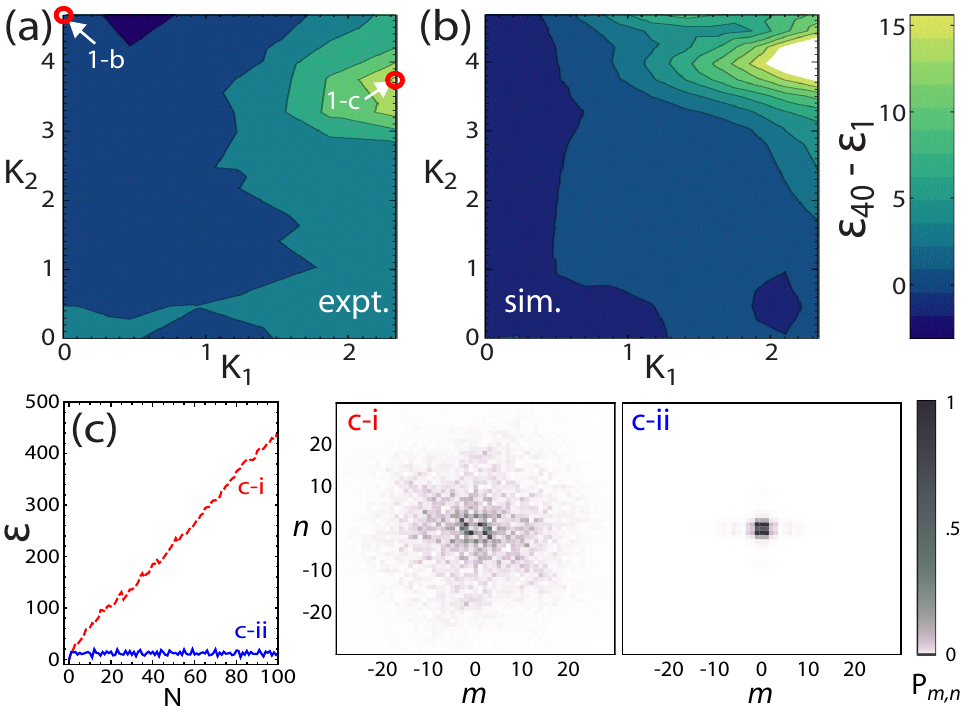}
\caption{
Delocalization of coupled kicked quantum rotors in two dimensions.
(a,b)~Measured and simulated change in $\varepsilon$ between $N=1$ to $40$ kicks as a function of $K_1$ and $K_2$ (8-9 sampled points in each direction), for off-resonant kicking as in Fig.~\ref{FIG:OffRes1}. The circles labeled 1-b and 1-c highlight data derived from the distributions shown in Figs.~\ref{FIG:Cartoon}~(b,c).
(c)~Simulated dependence of the energy $\varepsilon$ on $N$ for off-resonant $\delta$-kicking ($T = 36 \ \mu$s~;~$\tau = 10$~ns) by two equally deep lattices of $K_{1,2} = 5.8$, with (c-i, red dashed line) and without (c-ii, blue solid line) the cross-coupling term of $D_{m,n}$.
At right, 2D momentum distributions after $N=100$ kicks, decomposed with respect to the two lattices ($\mathrm{P}_{m,n} = |\langle m,n | \psi_N \rangle |^2$).
}
    \label{FIG:OffRes2}
\end{figure}

In contrast to the behavior seen for off-resonant kicking with a single lattice, we observe a delocalization of the atomic population into a nearly Gaussian momentum distribution when the second such lattice is added. This is shown in Fig.~\ref{FIG:Cartoon}~(c) for $K_{ \{1,2\} }= \tilde{K} = \{2.3,3.7\}$. For uniform kicking $K_1=K_2\equiv K_{1,2}$, we observe a threshold near $K_{1,2} \sim 2$  [cf. Fig.~\ref{FIG:OffRes1}~(b,c)], above which dynamical localization disappears and energy growth sets in. This finding is reproduced by the ground-state properties of the single-kick operator $\hat{U}$, for which a time-independent stroboscopic Floquet state analysis~\cite{Grempel-quantumNonintegrable-1984,Lemarie-AndersonTheoryExpt-2009} reveals a transition from localized to delocalized states at $K_{1,2} \sim 2.2$, cf. Fig.~\ref{FIG:OffRes1}~(d).

The full dependence of the observed growth on the kicking strengths $K_{1}$ and $K_2$ is shown in Fig.~\ref{FIG:OffRes2}~(a). We find essentially zero growth along the axes when either of the lattices are off-resonantly kicked alone, and a transition to diffusive growth when there is significant kicking by both lattices. The data are in good agreement with simulated quantum trajectories~\cite{DKR-SuppMatNEW}, cf. Fig.~\ref{FIG:OffRes2}~(b), including the position of maximal growth near $\tilde{K}$.
While the coupling term $\kappa \eta m n$ responsible for delocalization cannot be independently accessed in our experimental geometry, we can verify its role through numerical simulations as in Fig.~\ref{FIG:OffRes2}~(c), which can also probe more deeply into the diffusive, metallic phase. A linear energy growth, characteristic of diffusive transport, is clearly seen for strong kicking as a consequence of the coupling term, as opposed to a complete suppression of growth in its absence. The simulated growth is found to be very stable~\cite{DKR-SuppMatNEW}, persisting to timescales over 3 orders of magnitude longer than the system's quantum break time $t_B$ (limited only by the simulated system size). Furthermore, we find that in the limit of strong uniform driving the energy diffusion rate $\mathcal{D} = \Delta \varepsilon / \Delta N$ is not only non-zero, but asymptotically approaches the classical diffusion rate $\mathcal{D}_c = 2(1 + \eta^2)K_{1,2}^2/\kappa^2$~\cite{DKR-SuppMatNEW}.

We note that there exist fundamental differences between our observations and recent experimental investigations of the 3D Anderson model with cold atomic vapors driven at more than one frequency~\cite{Ringot-Delocalization-QuasiPeriodic,Chabe-AndersonMetal-2008,Lemarie-CriticalStateAnderson-2010}, where a metal to insulator transition results from the competition between disorder and tunneling.
Driving a single lattice at multiple temporal frequencies~\cite{Casati-3Danderson-1989,Tian-QuasiKickedAnderson-2011} is equivalent to a scenario of multiple uncoupled rotors. In contrast, we observe a transition in 2D that critically depends on the rotor-rotor coupling. The term $\kappa \eta m n$ in Eq.~\ref{EQ:Hamm2} represents a saddle potential which, when added to the two quadratic terms, breaks reflection symmetry about either of the two axes ($m \rightarrow -m$ and $n \rightarrow -n$), and on average breaks the $\mathbb{Z}_4$ rotational symmetry of the potential landscape. We point out that the observed coupling-induced diffusive behavior is particular to the pseudo-randomness of disorder in the $\delta$-KR system, and that it is not seen in simulations when purely random diagonal disorder is used instead. This is consistent with the fact that transitions from insulating to metallic behavior do not occur in 2D systems with purely random disorder, absent the breaking of time-reversal symmetry or spin-rotation invariance~\cite{Bellissard-2DelectronSO,Evan-Anderson-SO,Evers-AndTrans} (as due to strong magnetic fields or spin-orbit coupling in electronic systems).

\begin{figure}[t!]
    \centering
    \includegraphics[width=3.15in]{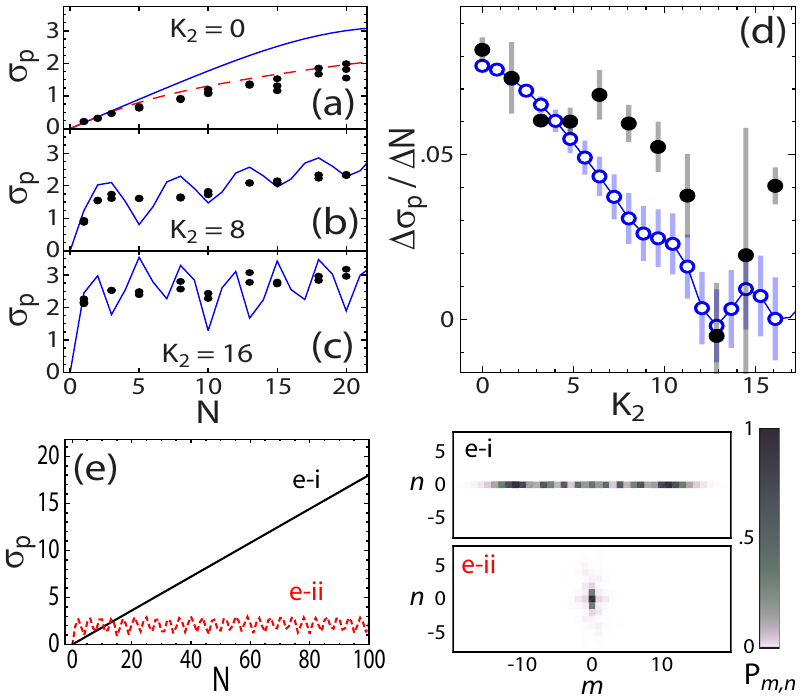}
\caption{Dynamical evolution of resonantly kicked matter waves ($K_1 = 1.6$) in the presence an additional off-resonant drive ($K_2$).
(a,b,c)~Momentum-width $\sigma_p$ of the atomic distribution as a function of kick number $N$, for a pulse period $T=124$~$\mu$s ($\kappa / 4\pi \approx 1$), and $K_2 = 0,8,16$.
Black points are data from individual experimental runs, the blue line is a numerical calculation for an initial plane-wave, while the dashed red curve takes into account finite-size corrections~\cite{DKR-SuppMatNEW}.
(d)~Growth rate of the momentum-width $\Delta \sigma_p / \Delta N$, determined by a linear fit to the $N$-dependence, as a function of $K_2$. The simulated growth rates (open blue circles) are scaled by a factor of $1/2$ to account for effects of finite size.\
(e)~Dependence of the momentum width $\sigma_p$ on kick number $N$, for the case of resonant kicking ($\kappa / 4\pi = 1$~;~$\tau = 10$~ns) with a single lattice ($K_{ \{1,2\} } = \{1.6,0\}$, black solid line e-i), and with an added deep incommensurate lattice ($K_{ \{1,2\} } = \{1.6,12.8\}$, red dashed line e-ii). Simulated characteristic momentum-space profiles in the 2D $m$-$n$ space are shown at right for $N=100$.
}
    \label{FIG:OnRes}
\end{figure}

Finally, we study our system at a quantum resonance of one of the rotors. For this purpose, we set $\kappa /4\pi \approx 1$ ($T=124$~$\mu$s) for the first lattice, while keeping the second lattice off-resonant ($\eta^2 \kappa /4\pi \approx 1.86$). When the second lattice is off, we observe a linear increase in the rms momentum width $\sigma_p$ (in units of $\hbar k_1$), cf. Fig.~\ref{FIG:OnRes}~(a). This is expected for constructive interference and characteristic of ballistic momentum-space transport. Adding the second lattice causes a reduction of the observed growth rate, depending on the strength of this lattice, cf.  Figs.~\ref{FIG:OnRes}~(b,c,d). To elucidate the mechanism behind this decrease, we show in Fig.~\ref{FIG:OnRes}~(e) simulations for a larger number of kicks (and for more ideal $\delta$-kicking exactly at resonance). The dynamics of $\sigma_p$ and the distributions of population within the 2D $m$-$n$ space suggest that the inhibition of ballistic transport is due not to a crossover to classical diffusion ($\sigma_p \propto \sqrt{N}$) as seen in the off-resonant case, but rather to the onset of dynamical localization in the strongly driven incommensurate lattice. The obvious underlying cause for the suppression is again the coupling term $\kappa \eta m n$, which destroys coherent phase revival between kicks. Remarkably, a complete suppression of resonant growth results even though the population spends nearly half of its time in the $n = 0$ subspace, where the coupling vanishes. This dynamical suppression of resonant quantum transport due to the admixture of off-resonant driving bears some resemblance to effects seen in other dynamical systems such as the Kapitza pendulum or ponderomotive potentials acting on charged particles.

To conclude, we have experimentally observed a quantum-to-classical transition from localized to delocalized dynamics in a system of coupled, kicked quantum rotors.
Further studies of our model system, as well as extensions to larger numbers of coupled rotors using several mutually incommensurate optical lattices, might help to provide insight into localization phenomena in nonlinear and disordered systems.

We thank Alexander Altland and Mikael Rechtsman for discussions, and Martin G.~Cohen and Thomas Bergeman for comments on the manuscript. This work was supported by NSF (PHY-0855643). B.G. and J.R. acknowledge support from the GAANN program of the US DoEd.


\begin{thebibliography}{35}%
\makeatletter
\providecommand \@ifxundefined [1]{%
 \@ifx{#1\undefined}
}%
\providecommand \@ifnum [1]{%
 \ifnum #1\expandafter \@firstoftwo
 \else \expandafter \@secondoftwo
 \fi
}%
\providecommand \@ifx [1]{%
 \ifx #1\expandafter \@firstoftwo
 \else \expandafter \@secondoftwo
 \fi
}%
\providecommand \natexlab [1]{#1}%
\providecommand \enquote  [1]{``#1''}%
\providecommand \bibnamefont  [1]{#1}%
\providecommand \bibfnamefont [1]{#1}%
\providecommand \citenamefont [1]{#1}%
\providecommand \href@noop [0]{\@secondoftwo}%
\providecommand \href [0]{\begingroup \@sanitize@url \@href}%
\providecommand \@href[1]{\@@startlink{#1}\@@href}%
\providecommand \@@href[1]{\endgroup#1\@@endlink}%
\providecommand \@sanitize@url [0]{\catcode `\\12\catcode `\$12\catcode
  `\&12\catcode `\#12\catcode `\^12\catcode `\_12\catcode `\%12\relax}%
\providecommand \@@startlink[1]{}%
\providecommand \@@endlink[0]{}%
\providecommand \url  [0]{\begingroup\@sanitize@url \@url }%
\providecommand \@url [1]{\endgroup\@href {#1}{\urlprefix }}%
\providecommand \urlprefix  [0]{URL }%
\providecommand \Eprint [0]{\href }%
\providecommand \doibase [0]{http://dx.doi.org/}%
\providecommand \selectlanguage [0]{\@gobble}%
\providecommand \bibinfo  [0]{\@secondoftwo}%
\providecommand \bibfield  [0]{\@secondoftwo}%
\providecommand \translation [1]{[#1]}%
\providecommand \BibitemOpen [0]{}%
\providecommand \bibitemStop [0]{}%
\providecommand \bibitemNoStop [0]{.\EOS\space}%
\providecommand \EOS [0]{\spacefactor3000\relax}%
\providecommand \BibitemShut  [1]{\csname bibitem#1\endcsname}%
\let\auto@bib@innerbib\@empty
\bibitem [{\citenamefont {Haake}(2010)}]{Haake}%
  \BibitemOpen
  \bibfield  {author} {\bibinfo {author} {\bibfnamefont {F.}~\bibnamefont
  {Haake}},\ }\href@noop {} {\emph {\bibinfo {title} {Quantum Signatures of
  Chaos}}},\ \bibinfo {edition} {3rd}\ ed.\ (\bibinfo  {publisher} {Springer},\
  \bibinfo {year} {2010})\BibitemShut {NoStop}%
\bibitem [{\citenamefont {Moore}\ \emph {et~al.}(1995)\citenamefont {Moore}
  \emph {et~al.}}]{Moore-Qdeltakicked-1995}%
  \BibitemOpen
  \bibfield  {author} {\bibinfo {author} {\bibfnamefont {F.~L.}\ \bibnamefont
  {Moore}} \emph {et~al.},\ }
  {\bibfield  {journal} {\bibinfo  {journal} {Phys. Rev. Lett.}\ }\textbf
  {\bibinfo {volume} {75}},\ \bibinfo {pages} {4598} (\bibinfo {year}
  {1995})}\BibitemShut {NoStop}%
\bibitem [{\citenamefont {Casati}\ \emph {et~al.}(1979)\citenamefont {Casati}
  \emph {et~al.}}]{Casati-Israilev-1979}%
  \BibitemOpen
  \bibfield  {author} {\bibinfo {author} {\bibfnamefont {G.}~\bibnamefont
  {Casati}} \emph {et~al.},\ }\enquote {\bibinfo {title} {Stochastic behavior
  in classical and quantum hamiltonian systems},}\ \ (\bibinfo  {publisher}
  {Springer, New York},\ \bibinfo {year} {1979})\ p.\ \bibinfo {pages}
  {334}\BibitemShut {NoStop}%
\bibitem [{\citenamefont {Hogg}\ and\ \citenamefont
  {Huberman}(1982)}]{HoggHuberman-Recurrences-1982}%
  \BibitemOpen
  \bibfield  {author} {\bibinfo {author} {\bibfnamefont {T.}~\bibnamefont
  {Hogg}}\ and\ \bibinfo {author} {\bibfnamefont {B.~A.}\ \bibnamefont
  {Huberman}},\ } {\bibfield
  {journal} {\bibinfo  {journal} {Phys. Rev. Lett.}\ }\textbf {\bibinfo
  {volume} {48}},\ \bibinfo {pages} {711} (\bibinfo {year} {1982})}\BibitemShut
  {NoStop}%
\bibitem [{\citenamefont {Fishman}\ \emph {et~al.}(1982)\citenamefont
  {Fishman}, \citenamefont {Grempel},\ and\ \citenamefont
  {Prange}}]{Fishman-Chaos-Recurr-Anderson-1982}%
  \BibitemOpen
  \bibfield  {author} {\bibinfo {author} {\bibfnamefont {S.}~\bibnamefont
  {Fishman}}, \bibinfo {author} {\bibfnamefont {D.~R.}\ \bibnamefont
  {Grempel}}, \ and\ \bibinfo {author} {\bibfnamefont {R.~E.}\ \bibnamefont
  {Prange}},\ } {\bibfield
  {journal} {\bibinfo  {journal} {Phys. Rev. Lett.}\ }\textbf {\bibinfo
  {volume} {49}},\ \bibinfo {pages} {509} (\bibinfo {year} {1982})}\BibitemShut
  {NoStop}%
\bibitem [{\citenamefont {Grempel}\ \emph {et~al.}(1984)\citenamefont
  {Grempel}, \citenamefont {Prange},\ and\ \citenamefont
  {Fishman}}]{Grempel-quantumNonintegrable-1984}%
  \BibitemOpen
  \bibfield  {author} {\bibinfo {author} {\bibfnamefont {D.~R.}\ \bibnamefont
  {Grempel}}, \bibinfo {author} {\bibfnamefont {R.~E.}\ \bibnamefont {Prange}},
  \ and\ \bibinfo {author} {\bibfnamefont {S.}~\bibnamefont {Fishman}},\ } {\bibfield  {journal} {\bibinfo
  {journal} {Phys. Rev. A}\ }\textbf {\bibinfo {volume} {29}},\ \bibinfo
  {pages} {1639} (\bibinfo {year} {1984})}\BibitemShut {NoStop}%
\bibitem [{\citenamefont {Ringot}\ \emph {et~al.}(2000)\citenamefont {Ringot}
  \emph {et~al.}}]{Ringot-Delocalization-QuasiPeriodic}%
  \BibitemOpen
  \bibfield  {author} {\bibinfo {author} {\bibfnamefont {J.}~\bibnamefont
  {Ringot}} \emph {et~al.},\ }
  {\bibfield  {journal} {\bibinfo  {journal} {Phys. Rev. Lett.}\ }\textbf
  {\bibinfo {volume} {85}},\ \bibinfo {pages} {2741} (\bibinfo {year}
  {2000})}\BibitemShut {NoStop}%
\bibitem [{\citenamefont {Chab\'e}\ \emph {et~al.}(2008)\citenamefont {Chab\'e}
  \emph {et~al.}}]{Chabe-AndersonMetal-2008}%
  \BibitemOpen
  \bibfield  {author} {\bibinfo {author} {\bibfnamefont {J.}~\bibnamefont
  {Chab\'e}} \emph {et~al.},\ }
  {\bibfield  {journal} {\bibinfo  {journal} {Phys. Rev. Lett.}\ }\textbf
  {\bibinfo {volume} {101}},\ \bibinfo {pages} {255702} (\bibinfo {year}
  {2008})}\BibitemShut {NoStop}%
\bibitem [{\citenamefont {Lemari\'e}\ \emph {et~al.}(2009)\citenamefont
  {Lemari\'e} \emph {et~al.}}]{Lemarie-AndersonTheoryExpt-2009}%
  \BibitemOpen
  \bibfield  {author} {\bibinfo {author} {\bibfnamefont {G.}~\bibnamefont
  {Lemari\'e}} \emph {et~al.},\ }
  {\bibfield  {journal} {\bibinfo  {journal} {Phys. Rev. A}\ }\textbf {\bibinfo
  {volume} {80}},\ \bibinfo {pages} {043626} (\bibinfo {year}
  {2009})}\BibitemShut {NoStop}%
\bibitem [{\citenamefont {Lemari\'e}\ \emph {et~al.}(2010)\citenamefont
  {Lemari\'e} \emph {et~al.}}]{Lemarie-CriticalStateAnderson-2010}%
  \BibitemOpen
  \bibfield  {author} {\bibinfo {author} {\bibfnamefont {G.}~\bibnamefont
  {Lemari\'e}} \emph {et~al.},\ } {\bibfield  {journal} {\bibinfo  {journal}
  {Phys. Rev. Lett.}\ }\textbf {\bibinfo {volume} {105}},\ \bibinfo {pages}
  {090601} (\bibinfo {year} {2010})}\BibitemShut {NoStop}%
\bibitem [{\citenamefont {Casati}\ \emph {et~al.}(1989)\citenamefont {Casati},
  \citenamefont {Guarneri},\ and\ \citenamefont
  {Shepelyansky}}]{Casati-3Danderson-1989}%
  \BibitemOpen
  \bibfield  {author} {\bibinfo {author} {\bibfnamefont {G.}~\bibnamefont
  {Casati}}, \bibinfo {author} {\bibfnamefont {I.}~\bibnamefont {Guarneri}}, \
  and\ \bibinfo {author} {\bibfnamefont {D.~L.}\ \bibnamefont {Shepelyansky}},\
  } {\bibfield  {journal} {\bibinfo
  {journal} {Phys. Rev. Lett.}\ }\textbf {\bibinfo {volume} {62}},\ \bibinfo
  {pages} {345} (\bibinfo {year} {1989})}\BibitemShut {NoStop}%
\bibitem [{\citenamefont {Ammann}\ \emph {et~al.}(1998)\citenamefont {Ammann},
  \citenamefont {Gray}, \citenamefont {Shvarchuck},\ and\ \citenamefont
  {Christensen}}]{Ammann-98}%
  \BibitemOpen
  \bibfield  {author} {\bibinfo {author} {\bibfnamefont {H.}~\bibnamefont
  {Ammann}}, \bibinfo {author} {\bibfnamefont {R.}~\bibnamefont {Gray}},
  \bibinfo {author} {\bibfnamefont {I.}~\bibnamefont {Shvarchuck}}, \ and\
  \bibinfo {author} {\bibfnamefont {N.}~\bibnamefont {Christensen}},\
  }\href@noop {} {\bibfield  {journal} {\bibinfo  {journal} {Phys. Rev. Lett.}\
  }\textbf {\bibinfo {volume} {80}},\ \bibinfo {pages} {4111} (\bibinfo {year}
  {1998})}\BibitemShut {NoStop}%
\bibitem [{\citenamefont {Klappauf}\ \emph {et~al.}(1998)\citenamefont
  {Klappauf} \emph {et~al.}}]{Klappauf-NoiseDiss-DynLoc-1998}%
  \BibitemOpen
  \bibfield  {author} {\bibinfo {author} {\bibfnamefont {B.~G.}\ \bibnamefont
  {Klappauf}} \emph {et~al.},\ }
  {\bibfield  {journal} {\bibinfo  {journal} {Phys. Rev. Lett.}\ }\textbf
  {\bibinfo {volume} {81}},\ \bibinfo {pages} {1203} (\bibinfo {year}
  {1998})}\BibitemShut {NoStop}%
\bibitem [{\citenamefont {d'Arcy}\ \emph
  {et~al.}(2001{\natexlab{a}})\citenamefont {d'Arcy} \emph
  {et~al.}}]{dArcy-QuantDiffusion-DKR-2001}%
  \BibitemOpen
  \bibfield  {author} {\bibinfo {author} {\bibfnamefont {M.~B.}\ \bibnamefont
  {d'Arcy}} \emph {et~al.},\ }
  {\bibfield  {journal} {\bibinfo  {journal} {Phys. Rev. Lett.}\ }\textbf
  {\bibinfo {volume} {87}},\ \bibinfo {pages} {074102} (\bibinfo {year}
  {2001}{\natexlab{a}})}\BibitemShut {NoStop}%
\bibitem [{\citenamefont {d'Arcy}\ \emph
  {et~al.}(2001{\natexlab{b}})\citenamefont {d'Arcy} \emph
  {et~al.}}]{dArcy-QaccelModes-Decoherence-2001}%
  \BibitemOpen
  \bibfield  {author} {\bibinfo {author} {\bibfnamefont {M.~B.}\ \bibnamefont
  {d'Arcy}} \emph {et~al.},\ }
  {\bibfield  {journal} {\bibinfo  {journal} {Phys. Rev. E}\ }\textbf {\bibinfo
  {volume} {64}},\ \bibinfo {pages} {056233} (\bibinfo {year}
  {2001}{\natexlab{b}})}\BibitemShut {NoStop}%
\bibitem [{\citenamefont {Zhang}\ \emph {et~al.}(2004)\citenamefont {Zhang}
  \emph {et~al.}}]{Zhang-Raizen-Interactions-2004}%
  \BibitemOpen
  \bibfield  {author} {\bibinfo {author} {\bibfnamefont {C.}~\bibnamefont
  {Zhang}} \emph {et~al.},\ }
  {\bibfield  {journal} {\bibinfo  {journal} {Phys. Rev. Lett.}\ }\textbf
  {\bibinfo {volume} {92}},\ \bibinfo {pages} {054101} (\bibinfo {year}
  {2004})}\BibitemShut {NoStop}%
\bibitem [{\citenamefont {Duffy}\ \emph {et~al.}(2004)\citenamefont {Duffy}
  \emph {et~al.}}]{Duffy-NonlinearDKR-2004}%
  \BibitemOpen
  \bibfield  {author} {\bibinfo {author} {\bibfnamefont {G.~J.}\ \bibnamefont
  {Duffy}} \emph {et~al.},\ }
  {\bibfield  {journal} {\bibinfo  {journal} {Phys. Rev. A}\ }\textbf {\bibinfo
  {volume} {70}},\ \bibinfo {pages} {041602} (\bibinfo {year}
  {2004})}\BibitemShut {NoStop}%
\bibitem [{\citenamefont {Shepelyansky}(1983)}]{Shepelyansky-Physica-1983}%
  \BibitemOpen
  \bibfield  {author} {\bibinfo {author} {\bibfnamefont {D.~L.}\ \bibnamefont
  {Shepelyansky}},\ }\href@noop {} {\bibfield  {journal} {\bibinfo  {journal}
  {Physica}\ }\textbf {\bibinfo {volume} {8D}},\ \bibinfo {pages} {208}
  (\bibinfo {year} {1983})}\BibitemShut {NoStop}%
\bibitem [{\citenamefont {Ott}\ \emph {et~al.}(1984)\citenamefont {Ott},
  \citenamefont {Antonsen},\ and\ \citenamefont
  {Hanson}}]{Ott-NoiseQuantChaos-1984}%
  \BibitemOpen
  \bibfield  {author} {\bibinfo {author} {\bibfnamefont {E.}~\bibnamefont
  {Ott}}, \bibinfo {author} {\bibfnamefont {T.~M.}\ \bibnamefont {Antonsen}}, \
  and\ \bibinfo {author} {\bibfnamefont {J.~D.}\ \bibnamefont {Hanson}},\
  } {\bibfield  {journal} {\bibinfo
   {journal} {Phys. Rev. Lett.}\ }\textbf {\bibinfo {volume} {53}},\ \bibinfo
  {pages} {2187} (\bibinfo {year} {1984})}\BibitemShut {NoStop}%
\bibitem [{\citenamefont {Graham}\ and\ \citenamefont
  {Kolovsky}(1996)}]{Kolovsky-2color-1996}%
  \BibitemOpen
  \bibfield  {author} {\bibinfo {author} {\bibfnamefont {R.}~\bibnamefont
  {Graham}}\ and\ \bibinfo {author} {\bibfnamefont {A.~R.}\ \bibnamefont
  {Kolovsky}},\ }\href@noop {} {\bibfield  {journal} {\bibinfo  {journal}
  {Physics Letters A}\ }\textbf {\bibinfo {volume} {222}},\ \bibinfo {pages}
  {47} (\bibinfo {year} {1996})}\BibitemShut {NoStop}%
\bibitem [{\citenamefont {Adachi}\ \emph {et~al.}(1988)\citenamefont {Adachi},
  \citenamefont {Toda},\ and\ \citenamefont
  {Ikeda}}]{Adachi-QuantToClass-CoupledRotors}%
  \BibitemOpen
  \bibfield  {author} {\bibinfo {author} {\bibfnamefont {S.}~\bibnamefont
  {Adachi}}, \bibinfo {author} {\bibfnamefont {M.}~\bibnamefont {Toda}}, \ and\
  \bibinfo {author} {\bibfnamefont {K.}~\bibnamefont {Ikeda}},\ } {\bibfield  {journal} {\bibinfo
  {journal} {Phys. Rev. Lett.}\ }\textbf {\bibinfo {volume} {61}},\ \bibinfo
  {pages} {659} (\bibinfo {year} {1988})}\BibitemShut {NoStop}%
\bibitem [{\citenamefont {Benvenuto}\ \emph {et~al.}(1991)\citenamefont
  {Benvenuto} \emph {et~al.}}]{Benvenuto-Nonlinear-1991}%
  \BibitemOpen
  \bibfield  {author} {\bibinfo {author} {\bibfnamefont {F.}~\bibnamefont
  {Benvenuto}} \emph {et~al.},\ }
  {\bibfield  {journal} {\bibinfo  {journal} {Phys. Rev. A}\ }\textbf {\bibinfo
  {volume} {44}},\ \bibinfo {pages} {R3423} (\bibinfo {year}
  {1991})}\BibitemShut {NoStop}%
\bibitem [{\citenamefont {Shepelyansky}(1993)}]{Shepel-Nonlinear-1993}%
  \BibitemOpen
  \bibfield  {author} {\bibinfo {author} {\bibfnamefont {D.~L.}\ \bibnamefont
  {Shepelyansky}},\ } {\bibfield
  {journal} {\bibinfo  {journal} {Phys. Rev. Lett.}\ }\textbf {\bibinfo
  {volume} {70}},\ \bibinfo {pages} {1787} (\bibinfo {year}
  {1993})}\BibitemShut {NoStop}%
\bibitem [{\citenamefont {Roati}\ \emph {et~al.}(2008)\citenamefont
  {Romanelli} \emph {et~al.}}]{RoatiIncLatt}%
  \BibitemOpen
  \bibfield  {author} {\bibinfo {author} {\bibfnamefont {G.}~\bibnamefont
  {Roati}} \emph {et~al.},\ }\href@noop {} {\bibfield  {journal} {\bibinfo
  {journal} {Nature}\ }\textbf {\bibinfo {volume} {453}},\ \bibinfo {pages}
  {895--898} (\bibinfo {year} {2008})}\BibitemShut {NoStop}%
\bibitem [{\citenamefont {Deng}\ \emph {et~al.}(1999)\citenamefont {Deng} \emph
  {et~al.}}]{Deng-Talbot-1999}%
  \BibitemOpen
  \bibfield  {author} {\bibinfo {author} {\bibfnamefont {L.}~\bibnamefont
  {Deng}} \emph {et~al.},\ }
  {\bibfield  {journal} {\bibinfo  {journal} {Phys. Rev. Lett.}\ }\textbf
  {\bibinfo {volume} {83}},\ \bibinfo {pages} {5407} (\bibinfo {year}
  {1999})}\BibitemShut {NoStop}%
\bibitem [{\citenamefont {Ryu}\ \emph {et~al.}(2006)\citenamefont {Ryu} \emph
  {et~al.}}]{Ryu-HighOrderRes-2006}%
  \BibitemOpen
  \bibfield  {author} {\bibinfo {author} {\bibfnamefont {C.}~\bibnamefont
  {Ryu}} \emph {et~al.},\ }
  {\bibfield  {journal} {\bibinfo  {journal} {Phys. Rev. Lett.}\ }\textbf
  {\bibinfo {volume} {96}},\ \bibinfo {pages} {160403} (\bibinfo {year}
  {2006})}\BibitemShut {NoStop}%
\bibitem [{\citenamefont {Romanelli}\ \emph {et~al.}(2005)\citenamefont
  {Romanelli} \emph {et~al.}}]{Romanelli-GeneralizeQRW-2005}%
  \BibitemOpen
  \bibfield  {author} {\bibinfo {author} {\bibfnamefont {A.}~\bibnamefont
  {Romanelli}} \emph {et~al.},\ }\href@noop {} {\bibfield  {journal} {\bibinfo
  {journal} {Physica A}\ }\textbf {\bibinfo {volume} {352}},\ \bibinfo {pages}
  {409} (\bibinfo {year} {2005})}\BibitemShut {NoStop}%
\bibitem [{\citenamefont {Romanelli}\ \emph {et~al.}(2007)\citenamefont
  {Romanelli}, \citenamefont {Siri},\ and\ \citenamefont
  {Micenmacher}}]{Romanelli-QKR-QRW-2007}%
  \BibitemOpen
  \bibfield  {author} {\bibinfo {author} {\bibfnamefont {A.}~\bibnamefont
  {Romanelli}}, \bibinfo {author} {\bibfnamefont {R.}~\bibnamefont {Siri}}, \
  and\ \bibinfo {author} {\bibfnamefont {V.}~\bibnamefont {Micenmacher}},\
  } {\bibfield  {journal} {\bibinfo
  {journal} {Phys. Rev. E}\ }\textbf {\bibinfo {volume} {76}},\ \bibinfo
  {pages} {037202} (\bibinfo {year} {2007})}\BibitemShut {NoStop}%
\bibitem [{\citenamefont {Abrahams}\ \emph {et~al.}(1979)\citenamefont
  {Abrahams}, \citenamefont {Anderson}, \citenamefont {Licciardello},\ and\
  \citenamefont {Ramakrishnan}}]{Abrahams-Anderson2D-1979}%
  \BibitemOpen
  \bibfield  {author} {\bibinfo {author} {\bibfnamefont {E.}~\bibnamefont
  {Abrahams}}, \bibinfo {author} {\bibfnamefont {P.~W.}\ \bibnamefont
  {Anderson}}, \bibinfo {author} {\bibfnamefont {D.~C.}\ \bibnamefont
  {Licciardello}}, \ and\ \bibinfo {author} {\bibfnamefont {T.~V.}\
  \bibnamefont {Ramakrishnan}},\ }\href@noop {} {\bibfield  {journal} {\bibinfo
   {journal} {Phys. Rev. Lett.}\ }\textbf {\bibinfo {volume} {42}},\ \bibinfo
  {pages} {673} (\bibinfo {year} {1979})}\BibitemShut {NoStop}%
\bibitem [{DKR()}]{DKR-SuppMatNEW}%
  \BibitemOpen
  \href@noop {} {}\bibinfo {note} {See Supplementary Material for more
  information on the theoretical analysis}\BibitemShut {NoStop}%
\bibitem [{\citenamefont {Geisel}\ \emph {et~al.}(1986)\citenamefont {Geisel},
  \citenamefont {Radons},\ and\ \citenamefont {Rubner}}]{Geisel-KAM-1986}%
  \BibitemOpen
  \bibfield  {author} {\bibinfo {author} {\bibfnamefont {T.}~\bibnamefont
  {Geisel}}, \bibinfo {author} {\bibfnamefont {G.}~\bibnamefont {Radons}}, \
  and\ \bibinfo {author} {\bibfnamefont {J.}~\bibnamefont {Rubner}},\ } {\bibfield  {journal} {\bibinfo
  {journal} {Phys. Rev. Lett.}\ }\textbf {\bibinfo {volume} {57}},\ \bibinfo
  {pages} {2883} (\bibinfo {year} {1986})}\BibitemShut {NoStop}%
\bibitem [{\citenamefont {Tian}\ \emph {et~al.}(2011)\citenamefont {Tian},
  \citenamefont {Altland},\ and\ \citenamefont
  {Garst}}]{Tian-QuasiKickedAnderson-2011}%
  \BibitemOpen
  \bibfield  {author} {\bibinfo {author} {\bibfnamefont {C.}~\bibnamefont
  {Tian}}, \bibinfo {author} {\bibfnamefont {A.}~\bibnamefont {Altland}}, \
  and\ \bibinfo {author} {\bibfnamefont {M.}~\bibnamefont {Garst}},\
  }\href@noop {} {\bibfield  {journal} {\bibinfo  {journal} {Phys. Rev. Lett.}\
  }\textbf {\bibinfo {volume} {107}},\ \bibinfo {pages} {074101} (\bibinfo
  {year} {2011})}\BibitemShut {NoStop}%
\bibitem [{\citenamefont {Bellissard}\ \emph {et~al.}(1986)\citenamefont
  {Bellissard}, \citenamefont {Grempel}, \citenamefont {Martinelli},\ and\
  \citenamefont {Scoppola}}]{Bellissard-2DelectronSO}%
  \BibitemOpen
  \bibfield  {author} {\bibinfo {author} {\bibfnamefont {J.}~\bibnamefont
  {Bellissard}}, \bibinfo {author} {\bibfnamefont {D.~R.}\ \bibnamefont
  {Grempel}}, \bibinfo {author} {\bibfnamefont {F.}~\bibnamefont {Martinelli}},
  \ and\ \bibinfo {author} {\bibfnamefont {E.}~\bibnamefont {Scoppola}},\
  }\href@noop {} {\bibfield  {journal} {\bibinfo  {journal} {Phys. Rev. B}\
  }\textbf {\bibinfo {volume} {33}},\ \bibinfo {pages} {641} (\bibinfo {year}
  {1986})}\BibitemShut {NoStop}%
\bibitem [{\citenamefont {Evangelou}(1995)}]{Evan-Anderson-SO}%
  \BibitemOpen
  \bibfield  {author} {\bibinfo {author} {\bibfnamefont {S.~N.}\ \bibnamefont
  {Evangelou}},\ }\href@noop {} {\bibfield  {journal} {\bibinfo  {journal}
  {Phys. Rev. Lett.}\ }\textbf {\bibinfo {volume} {75}},\ \bibinfo {pages}
  {2550} (\bibinfo {year} {1995})}\BibitemShut {NoStop}%
\bibitem [{\citenamefont {Evers}\ and\ \citenamefont
  {Mirlin}(2008)}]{Evers-AndTrans}%
  \BibitemOpen
  \bibfield  {author} {\bibinfo {author} {\bibfnamefont {F.}~\bibnamefont
  {Evers}}\ and\ \bibinfo {author} {\bibfnamefont {A.~D.}\ \bibnamefont
  {Mirlin}},\ }\href@noop {} {\bibfield  {journal} {\bibinfo  {journal} {Rev.
  Mod. Phys.}\ }\textbf {\bibinfo {volume} {80}},\ \bibinfo {pages} {1355}
  (\bibinfo {year} {2008})}\BibitemShut {NoStop}%
\end{thebibliography}
%

\clearpage
\newpage


\section*{Supplementary material}

# Supplementary Material for: "Evidence for a Quantum-to-Classical Transition in a Pair of Coupled Quantum Rotors"

Bryce Gadway,* Jeremy Reeves, Ludwig Krinner, and Dominik Schneble

*Department of Physics and Astronomy, Stony Brook University, Stony Brook, NY 11794-3800, USA*

(Dated: May 2, 2013)

## SIMULATED QUANTUM DYNAMICS

We compare our experimental data to simulated quantum trajectories of observables such as the atoms' mean energy $\varepsilon$ and rms momentum width $\sigma_p$. For the numerical trajectories, we consider the basis of plane-wave states coupled by the two light fields $|m,n\rangle$, with momentum $p_{m,n}/\hbar k_1 = 2m + 2n\eta$. In determining $\hat{U}$, we do not consider the limit of $\delta$-like pulses, but take into account effects of finite pulse duration. Here, $\hat{U} = \hat{R}^{-1}\hat{E}_L\hat{R}\hat{E}_0$, where $\hat{E}_0$ accounts for free evolution for a time $T - \tau$, $\hat{E}_L$ describes evolution of the lattice eigenstates (as determined by numerical diagonalization) during the pulse for a time $\tau$, and $\hat{R}$ maps the free-particle eigenstates onto those of the lattice. After a train of $N$ pulses, the atomic wavefunction is then given by $|\psi_N\rangle = \hat{U}^N|\psi_0\rangle$, with $|\psi_0\rangle = |0,0\rangle$ the initial state. While we typically consider up to $\pm 15$-$20$ momentum orders for each lattice, we restrict the determination of $\varepsilon$ and $\sigma_p$ to $|p|/\hbar k_1 \leq 15$, to agree with the range of experimentally observed momenta.

## SIMULATED CLASSICAL DYNAMICS

It is well known that classical localization can occur in kicked-rotor systems for small values of the stochasticity parameter, where the classical system is not globally chaotic but supports bounded orbits due to Kolmogorov–Arnol'd–Moser (KAM) barriers [1]. To demonstrate that this effect does not contribute to the observed localization in momentum space in Fig. 2 (a,c,d) of the main text (i.e. for atoms kicked by only a single optical lattice), we compare the experimental data and the simulated quantum trajectories of the per-particle energy $\varepsilon$ as a function of kick number $N$ to simulated classical dynamics, averaged over $10^5$ classical trajectories. To briefly detail the determination of the averaged classical dynamics, each individual classical trajectory is determined by probabilistically projecting out the particle's momentum space wavefunction after each application of the kick operator $\hat{U}$. We show in FIG. 1 (i-iv) the classical dynamics of $\varepsilon$, as averaged over $n = 10$, $10^2$, $10^3$, and $10^5$ individual trajectories, for the same system parameters as used in Fig. 2 (a) of the main text [$T = 36$ $\mu$s ; $\tau = 2$ $\mu$s ; $K_2 = 4.6$ ($s_2 = 100$) ; $K_1 = 0$]. For large sample sizes, the classical growth rate is found to be in fair agreement with the expected classical diffusion constant [2] in the applied second lattice,

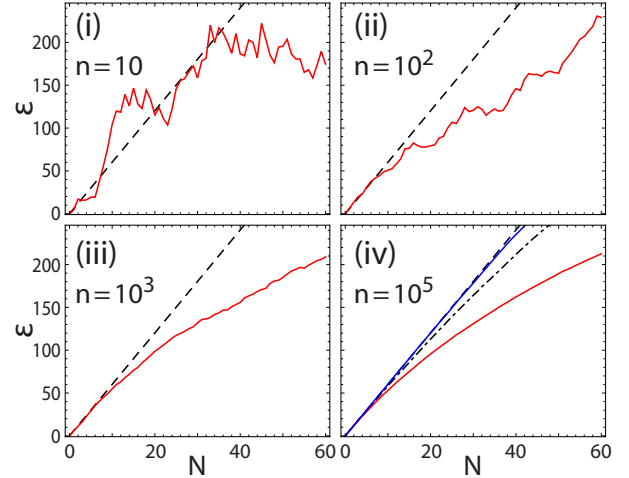

FIG. 1. Averaged classical dynamics of the energy $\varepsilon$ as a function of kick number $N$ for different sample sizes consisting of $n$ trajectories, for off-resonant kicking with the second lattice [$K_2 = 4.6$ ($s_2 = 100$) ; $K_1 = 0$ ; $\tau = 2$ $\mu$s]. (i-iv) Averaged trajectories are shown for the cases of $n = 10$, $10^2$, $10^3$, and $10^5$ as solid red lines. In all plots, we also show as a dashed black line the expected classical diffusion for delta-function kicks. In (iv), we also show classical trajectories for the cases in which the pulse area is the same, but with smaller pulse durations of $\tau = 1$ $\mu$s (black dashed-dotted line) and 0.1 $\mu$s (blue solid line).

$\mathcal{D} = \Delta\varepsilon/\Delta N \sim \eta^2 V_2^2 \tau^2/2\hbar^2 = 2\eta^2 K_2^2/\kappa^2$. For large kick numbers, the simulated classical trajectories are seen to grow more slowly than expected from the classical diffusion constant. This deviation is due to the finite kick duration ($\tau = 2$ $\mu$s) [3]. In FIG. 1 (iv) we also plot the simulated trajectories for the same pulse area but for $\tau = 1$ $\mu$s and 0.1 $\mu$s, showing that better agreement is reached in the limit of $\delta$-like pulses.

## TIME-INDEPENDENT ANALYSIS OF KICK OPERATOR $\hat{U}$

In the experiment, we are restricted to a modest number of kicks to maintain a near-field treatment (assuming identical, overlapped spatial wavefunctions for all the matter-wave fields) and to minimize contributions from nonlinear atom-atom interactions. To compare to the expectation for long-term behavior, we also study the time-independent properties of the Floquet (or Bloch–Floquet) quasi-energy eigenstates of the kick operator $\hat{U}$.

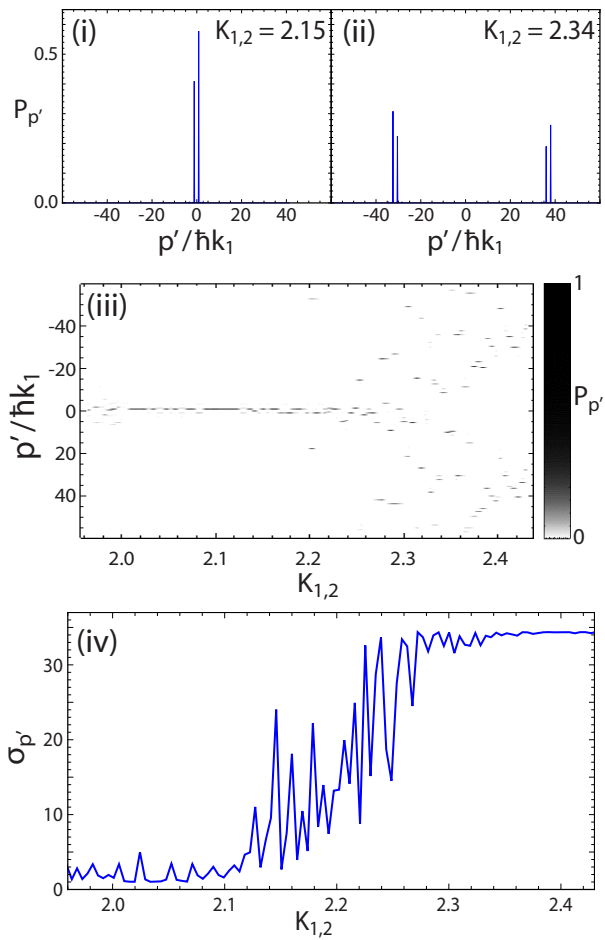

FIG. 2. Time-independent analysis of the kick operator $\hat{U}$ for off-resonant kicking ($T = 36$ μs) by both lattices, with equal kicking strengths $K_{1,2}$ for each. (i,ii) The momentum distribution of the lowest energy Floquet eigenstate of $\hat{U}$, $|\psi_0^U\rangle$, in the plane-wave basis. The momenta $p$ are shifted by the mean value $\bar{p} = \langle\psi_0^U|\hat{p}|\psi_0^U\rangle$ to $p' = p - \bar{p}$. For low values of the kicking strength, as in (i), the eigenstates are localized in momentum space, while for larger values they became delocalized as in (ii). (iii) Momentum distributions as in (i,ii) as a function of the stochasticity parameter $K_{1,2}$. Beyond a value of $K_{1,2} \sim 2.2$ a bifurcation and delocalization of the momentum distributions is observed. (iv) For the same range of stochasticity parameter values as in (iii), we plot the rms momentum width $\sigma_{p'}$ of the lowest energy eigenstates of $\hat{U}$, which exhibits a sharp increase across a value $K_{1,2} \sim 2.2$.

For regimes in which the growth dynamics result in dynamical localization, these Floquet eigenstates should all be localized in momentum space, while they are delocalized in the case of diffusive growth.

We first investigate the case of kicking off-resonantly with two lattices of equal strength, using the same parameter values as in Fig. 2 (e) of the main text, with variable stochasticity parameter $K_1 = K_2 \equiv K_{1,2}$. In Fig. 2 (i,ii), we plot the momentum space distribution of the lowest energy Floquet eigenstates of $\hat{U}$, $|\psi_0^U\rangle$, for the values $K_{1,2} = 2.15$ and $2.34$. The plotted distributions are shifted by the mean momentum $\bar{p} = \langle\psi_0^U|\hat{p}|\psi_0^U\rangle$ to $p' = p - \bar{p}$. While the distribution is localized for the weaker kicking strength, it is delocalized into two regions for the larger strength. In fact, in Fig. 2 (iii) we observe a bifurcation of the eigenstate distributions as the stochasticity parameter $K_{1,2}$ is increased. We can readily calculate the rms momentum width $\sigma_{p'}$ of these distributions, and as shown in Fig. 2 (iv) there is a sharp rise in the width across a value of $K_{1,2} \sim 2.2$. Not surprisingly, this value is in general agreement with the observed onset of non-zero energy growth versus kick number in Fig. 2 (e) of the main text.

In addition, we perform a Floquet eigenstate analysis in the full parameter space of $K_1$ and $K_2$. In Fig. 3 (i), we plot as white the regions in which the Floquet eigenstates are localized in momentum space and as dark blue the regions in which they are delocalized, with the criterion for delocalization being that more than 10% of the population resides further than 5.5 momentum units ($\hbar k_1$) away from the most populated mode. This plot shows localization along either axis, where dynamical localization is expected for any value of the kicking strength when only a single lattice is used, while delocalization generally occurs beyond some line in the parameter space spanned by $K_1$ and $K_2$ when the strengths of each of the incommensurate frequency components are significant. We also show in Fig. 3 (ii) the simulated change in $\varepsilon$ between one and $N = 40$ kicks, as in Fig. 3 (b) of the main text but over a larger range of $K_1$ values. This plot demonstrates that qualitatively similar regions of localization and delocalization are also observed in the energy growth dynamics.

## LONG-TIME GROWTH DYNAMICS AND CLASSICAL DIFFUSION

As discussed in the main text, we find that the observed diffusive behavior is extremely stable, and persists on timescales up to 3 orders of magnitude longer than the "quantum break time" $t_B$ in our system (limited only by the simulated system size). To observe growth at these very long times, when population reaches much higher momentum orders, we increase the size of our simulations to include up to ±48 momentum orders for each lattice, i.e. roughly 9400 $|m, n\rangle$ states in total. In addition to allowing for the investigation of growth at long times, the inclusion of a much larger system size allows us to investigate growth for very larger kicking strengths. We find that for increasing uniform kicking strength $K_1 = K_2 \equiv K_{1,2}$, the rate of diffusive energy growth $\mathcal{D} = \Delta\varepsilon/\Delta N$ first becomes non-zero at $K_{1,2} \approx 2$ (as also found in Fig. 2 (c,d) of the main text), and then increases monotonically with the kicking strength, as one may expect. More surprising is that in the limit of very

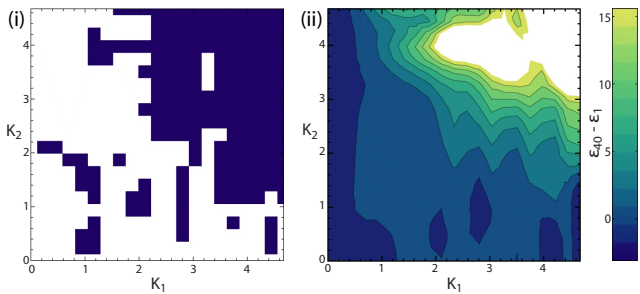

FIG. 3. Localization-delocalization transition in the $K_1$-$K_2$ plane. (i) As a function of the stochasticity parameters $K_1$ and $K_2$, we plot as white the regions in which the Floquet eigenstates of $\hat{U}$ are localized in momentum space and as blue the regions in which they are delocalized, as detailed in the text. A general trend of delocalization for strong kicking with both lattices, and localization when kicking with only a single lattice, is observed. (ii) Here we replot the simulation data from Fig. 3 (b) of the main text, showing the change in energy $\varepsilon$ from the first to $N = 40$ kicks, but over a larger range of $K_1$ values. Behavior qualitatively similar to the localization-delocalization plot in (i) can be observed in this dynamical response data.

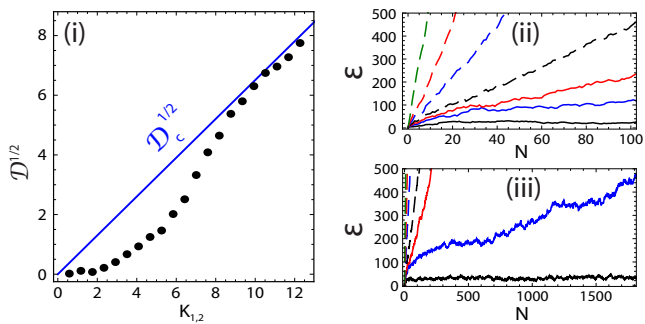

FIG. 4. Long-time dynamics and classical energy diffusion. (i) The energy diffusion rate $\mathcal{D} = \Delta\varepsilon/\Delta N$, as fit over the first 50 kicks, as a function of the uniform kicking strength $K_{1,2}$, with inter-rotor coupling present. The blue line represents the classical diffusion rate for our system parameters. (ii) The observed growth in $\varepsilon$ vs. kick number for several kicking strengths (in ascending order: $K_{1,2} = \{3.51, 4.68, 5.26, 5.84, 7.01, 8.18, 11.69\}$). (iii) For the same simulated trajectories as in (ii), we highlight the extremely long timescales over which growth can be observed.

strong kicking the energy diffusion rate approaches the classical value $\mathcal{D}_c = 2(1+\eta^2)K_{1,2}^2/\kappa^2$ (for uncorrelated driving by our two lattices). The long-time growth dynamics, as well as the approach to diffusive energy growth at the classical rate, are shown in Fig. 4.

## FINITE SIZE EFFECTS

In the case of driving on-resonance with only a single lattice, as studied in Fig. 4 (a) of the main text, significant deviations from the plane-wave simulations occur owing to finite-size effects. The observed growth rate is lower by a factor of $\sim 2$ from that of the plane-wave simulations, due to the finite size ($\sim 18~\mu$m) of the BEC along $z$, which gives rise to an initial rms momentum width $\Delta p \approx 0.03\hbar k_1$.

We assume a Gaussian momentum distribution with width $\Delta p = 0.03\hbar k_1$, and we employ the analytical formula [4, 5] describing the mean energy (in units of $E_R$) after $N$ lattice pulses (duration $\tau$, with a lattice of depth $s_1 E_R$) for particles with initial quasimomentum $\beta = p/2\hbar k_1$,

$$\varepsilon(\beta, N) = \frac{s_1^2 E_R^2 \tau^2}{2\hbar^2} \left[\frac{\sin^2(2\pi\beta N)}{\sin^2(2\pi\beta)}\right]. \quad (1)$$

This calculation results in the red dashed line of Fig. 4 (a), which agrees much better with the experimental data than does the plane-wave simulation. This simple analytical form relates directly only to the case of single-lattice kicking, and as such is not applied to the two-lattice cases of Figs. 4 (b,c). However, as the effects of finite-size effects are most significant when there is constructive interference at resonance, the cases of two-lattice kicking, where resonant growth is additionally suppressed by dynamical localization, are in better agreement with the plane-wave simulations.

[*] present address: JILA, University of Colorado, 440 UCB, Boulder, CO 80309-0440
[1] D. R. Grempel, R. E. Prange, and S. Fishman, Phys. Rev. A **29**, 1639 (1984).
[2] G. Lemarié et al., Phys. Rev. A **80**, 043626 (2009).
[3] B. Gadway et al., Opt. Express **17**, 19173 (2009).
[4] S. Wimberger et al., Nonlinearity **16**, 1381 (2003).
[5] C. Ryu et al., Phys. Rev. Lett. **96**, 160403 (2006).



\end{document}